\documentclass[vecphys]{svmult}

\usepackage{makeidx}         % allows index generation
\usepackage{graphicx}        % standard LaTeX graphics tool
\usepackage{multicol}        % used for the two-column index
\usepackage{cite}            % adjusts the "syntax" of the refs in the
\usepackage[bottom]{footmisc}% places footnotes at page bottom
\usepackage[english]{babel}
\usepackage{graphicx}
\usepackage{verbatim}
\makeindex             % used for the subject index

\def \mb{\begin{displaymath}} 
\def \me{\end{displaymath}} 
\def \eb{\begin{equation}} 
\def \ee{\end{equation}}   
\def\expect#1{\mathinner{\langle{#1}\rangle}}

{\catcode`\|=\active 
  \gdef\expect#1{\left<\mathcode`\|"8000\let|\bravert {#1}\right>}}
\def\bravert{\egroup\,\vrule\,\bgroup}

\begin{document}

\title{Electron transport through molecules in the Kondo regime: the
  role of molecular vibrations}
\titlerunning{Phonons and Kondo effect}
\author{J. Mravlje$^{1}$ \and A. Ram\v{s}ak$^{2,1}$}
\institute{Jo\v{z}ef Stefan Institute, Ljubljana, Slovenia, 
\texttt{jernej.mravlje@ijs.si} \and Faculty of Mathematics and Physics, University of Ljubljana,
Slovenia}

\maketitle

\begin{abstract}
  We discuss the electronic transport through molecules in the Kondo
  regime. We concentrate here on the influence of molecular vibrations.  Two
  types of vibrations are investigated: (i) the breathing internal
  molecular modes, where the coupling occurs between the molecular
  deformation and the charge density, and (ii) the oscillations of
  molecule between the contacts, where the displacement 
  affects the tunneling. The system is described by models
  which are solved numerically using Sch\"onhammer-Gunnarsson's
  projection operators and Wilson's numerical renormalization group
  methods.

  Case (i) is considered within the Anderson-Holstein model. Here the
  influence of the phonons is mainly to suppress the repulsion between
  the electrons at the molecular orbital.  Case (ii) is described by a
  two-channel Anderson model with phonon-assisted hybridization.
  In both cases, the coupling to electrons softens the vibrational mode
  and in the strong coupling regime makes the displacement unstable to
  perturbations that break the symmetry of the confining
  potential. For instance, in case (ii) when the frequency of
  oscillations decreases below the magnitude of perturbation breaking
  the left-right symmetry, the molecule will be abruptly attracted to
  one of the electrodes. In this regime, the Kondo temperature
  increases but the conductance through the molecule is suppressed.

\end{abstract}

% \pacs{72.15.Qm,73.23.-b,73.22.-f}
%71.27.+a 	Strongly correlated electron systems; heavy fermions
%73.23.-b 	Electronic transport in mesoscopic systems
%73.23.Hk 	Coulomb blockade; single-electron tunneling
%72.15.Qm 	Scattering mechanisms and Kondo effect (see also
%               75.20.Hr local moments in compounds and alloys; Kondo effect, valence
%               fluctuations, heavy fermions in magnetic properties and materials) 
%73.22.-f 	Electronic structure of nanoscale materials: clusters,
%               nanoparticles, nanotubes, and nanocrystals

\section{INTRODUCTION}
The fast pace of the computer industry is mainly driven by the
miniaturization of elements in microprocessors. The ultimate limit of
the miniaturization is to control the current through individual
molecules and it is remarkable that transistors based on single
molecules bridging metallic electrodes have already been produced and
their current-voltage characteristics have been measured \cite{park02,
  nitzan03,tao06,galperin07}. Such molecular junctions are produced
using mechanical breaking or electromigration tecniques which
currently do not allow for scaling up to larger circuits, but they
already provide information on the electron transport on the nanoscale
that could be essential to the circuitry of tomorrow
\cite{natelson09}.

Moreover, because the transmission through a molecule is sensitive to
its immediate electro-chemical (and also magnetic) environment such
devices could work as molecular sensors. For instance, the binding of
guest species to a single host molecule bridging two electrodes has
already been discerned in conductance measurements \cite{xiao04}. The
studies of conductance could thus enable recognition of single
molecules and thereby realize the ultimate limit of analytical
chemistry.

In certain regimes the molecular junctions exhibit the Kondo effect
\cite{hewson_book, park02,yu04_2,pasupathy05}: the anomalous behavior
of conductance due to the increased scattering rate driven by the
residual spin (i.e., the quantum impurity) localized at the molecular
orbital. The molecular transistors thus provide the nanoscopic
realization of quantum impurity models and can be used thus also as a
laboratory to investigate many-particle physics, for instance the
quantum phase transitions \cite{roch08}.

The transport through molecules is affected by molecular vibrations (MV).
The molecular internal vibrational modes and oscillations of molecules
with respect to the electrodes explain the side-peaks observed in the
non-linear conductance \cite{park02, yu04_2,pasupathy05, zhitenev02}.
In this article we are interested in the effects of coupling to the
MV at low temperatures and at a small bias
(i.e. in the Kondo regime) and their signature in the dependence of
the conductance on the gate-voltage. 

We consider two different types of molecular vibrations. (i) In the
case of breathing molecular modes, i.e., when the MV couple to the
electron density, the electron-electron repulsion is effectively
diminished and the electron effective mass is enhanced. (ii) When the
molecule itself oscillates between the contacts, i.e., the MV modulate
the tunneling, the effective repulsion is unmodified but the
asymetrical part of the modulation introduces the charge fluctuations
in the odd conduction channel, which leads to the competition between
odd and even channel that can result (albeit with some articial
fine-tuning of the models, as we discuss) in the ground state with the
non-Fermi liquid 2 channel Kondo fixed point.

Although the influence of the MV on the electrons differs profoundly
in these two cases, the back-action of electrons to molecular
vibrations is universal. The coupling to electrons tends to soften the
molecular modes (diminish their effective frequencies). This softening
is related to the increased charge-susceptibility in case (i) or
increased susceptibility to breaking of inversion symmetry in case (ii).
The result is a suppressed conductance with simultaneous increase
of the Kondo temperature.

This contribution provides an overview of our work on quantum impurity
models coupled to phonons \cite{mravlje05,mravlje06, mravlje07,
  mravlje08}. Because of lack of space we here develop only the main
ideas and refer the reader to these articles and the references
therein.  For background on the Anderson-Holstein model we
specifically refer also to \cite{hewson02, cornaglia04, cornaglia05a}
and for the work on the oscillating molecules we refer also to
\cite{balseiro06, lucignano08}.

\section{Coupling of vibration to charge: Anderson-Holstein model}
\label{sec:lang}
\begin{figure}[h]
\begin{center}\includegraphics[%
  width=45mm,
  keepaspectratio]{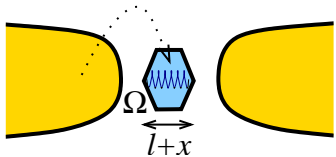}\end{center}  % thesis/Fig2.eps
\caption{\label{cap:Fig0} Transport through a molecule with a
  breathing mode. }
\end{figure}

Consider a molecule with a breathing mode,
trapped between two electrodes as depicted schematically on
Fig.~\ref{cap:Fig0}. Assuming that a single molecular orbital is
relevant for the electron transport (experimentally this assumption is
supported by wide inter-orbital energy spacings \cite{yu04_2}), the
system can be described by the Anderson-Holstein Hamiltonian,
\begin{equation} 
H=\sum_{k\alpha} \epsilon_k n_{k\alpha} +
  \sum_{k\alpha\sigma}\left( V_{k\alpha} c_{k\alpha\sigma}^\dagger d_\sigma + h.c. \right)
  + \epsilon n + Un_\uparrow n_\downarrow + M (n-1)x + \Omega
  a^\dagger a
\label{eq:hami}
\end{equation}
describing bands of noninteracting electrons in the left ($\alpha=L$)
and right ($\alpha=R$) electrodes, with energies $\epsilon_k$,
$n_{k\alpha}=n_{k\alpha\uparrow} + n_{k\alpha\downarrow}$, which are
counted by $n_{k\alpha \sigma}=c_{k\alpha\sigma}^\dagger c_{k\alpha
  \sigma}$. Likewise $n=n_\uparrow+n_\downarrow$ with
$n_\sigma=d^\dagger_\sigma d_\sigma$ counts the electrons at the
molecular orbital with the single-electron energy denoted by
$\epsilon$; $c_{k\sigma},d_\sigma$ are the electron anhilation and
$c_{k\sigma}^\dagger, d_\sigma^\dagger$ the electron creation
operators. The tunneling matrix element between $k$-state in the
electrode $\alpha$ and the molecular orbital is given by
$V_{k\alpha}$. The electrons in the electrodes are assumed
noninteracting, the electron repulsion between electrons at the
molecular orbital is $U$. The charge on the molecular orbital couples
to the displacement of the phonon mode $x=a+a^\dagger$
[$a^{(\dagger)}$ is the phonon anhilation (creation) operator] via a
Holstein coupling of strength $M$, while the frequency of the internal
vibrational mode of isolated molecule is $\Omega$.

Assuming (for simplicity) that the system is inversion symmetric
(meaning that the tunneling to the left is equal to the tunneling to
the right electrode, $V_{kL} =V_{kR}$) it is convenient to define
operators in the electrodes which are even/odd upon inversion
\begin{equation}
c_{e,o} = \frac{1}{\sqrt{2}} \left(c_L\pm c_R\right). 
\end{equation}
By rewriting the Hamiltonian in the new basis, likewise, the coupling
to even channel is given by $V_e= (1/\sqrt{2}) (V_L+V_R)$ and the
coupling to the odd channel vanishes $V_o=(1/\sqrt{2}) (V_L-V_R)=0$.
 It is thus sufficient to retain only the even operators explicitly and
describe the system by a single channel Anderson-Holstein model.
\begin{equation} 
  H=\sum_k \epsilon_k n_{ke} + V \sum_\sigma \left( f_\sigma^\dagger d_\sigma
 + h.c.  \right)+ \epsilon n + Un_\uparrow n_\downarrow + M (n-1)x + \Omega
  a^\dagger a ,  
\end{equation}
where $f_\sigma$ is the linear combination of the conduction electrons
to which the molecular orbital (i.e., the impurity) couples directly,
\begin{equation} f_\sigma=(\sum_k V_{ke} c_{ke \sigma})/(\sum_k
|V_{ke}|^2)^{1/2}=(1/\sqrt{N}) \sum_k c_{ke \sigma}.  \label{eq:fs}
\end{equation}

\subsection{Analytical considerations}

A convenient starting point for the analysis of the model is to
perform the unitary displaced oscillator transformation. One obtains:
\begin{eqnarray}
  H' &= &e^{\frac{M}{\Omega} (a-a^\dagger) (n-1)} H
  e^{-\frac{M}{\Omega}(a-a^\dagger)(n-1)}=  \label{eq:langfirs}   \\
&=& \epsilon_{\mathrm{eff}} n + U_{\mathrm{eff}}  
  n_\uparrow n_\downarrow +  V\left[\sum_{ \sigma}  f_{\sigma}^\dagger d'_\sigma
    +h.c. \right] + \sum_k
  \epsilon_k n_k + \Omega a^\dagger a, \nonumber
\end{eqnarray}
where
\begin{equation}
d'=e^{-\frac{M}{\Omega}(a-a^\dagger)} d; \; U_\mathrm{eff}=U-2M^2/\Omega;
\; \epsilon_{\mathrm{eff}} =\epsilon + M^2/\Omega \nonumber. 
\end{equation}
The transformed Hamiltonian $H'$ is of the same form as $H$, but with
a 
reduced repulsion $U_\mathrm{eff}$. The coupling to the phonons is
hidden in the transformed operator $d'$. The transformed boson
operators read,
\begin{equation}
a'=a-\frac{M}{\Omega} (n-1).
\end{equation}
The displacement  is shifted 
depending on the occupancy of the molecular orbital
\begin{equation}
x'=x-2M (n-1)/\Omega.
\end{equation}
There is an interesting large-frequency limit $M/\Omega \to 0$, but
$M^2/\Omega$ finite, where the coupling to phonons is entirely
described in terms of the effective parameters of the model.

The same result follows from considering the EP
interaction perturbatively \cite{hewson02}. A pair of EP
vertices and a phonon propagator can be formally substituted by a
frequency-dependent point electron-electron interaction vertex,
The influence of the phonon mode is then retained in
the frequency dependence of the interaction
\begin{equation}
  U(\omega) = U + M^2 D_0(\omega) =U -{{2M^2 \Omega} \over {\Omega^2 -\omega^2}},
\end{equation}
because $D_0(\omega)= 2\Omega/(\omega^2-\Omega^2)$.  At low
frequencies the interaction is screened due to the formation of
bipolarons $U(\omega\to 0)=U_\mathrm{eff}$, at high frequencies the
bare interaction is recovered. If $\Omega$ is large, then the
low-energy behavior is given entirely by the Anderson model with
$U_\mathrm{eff}$. % The numerical results valid for this regime are
% presented on Fig.~\ref{ah_u}.
Note that effective repulsion can become
negative; that provides the motivation for studies in the $U<0$
regime.

\subsection{Numerical results}
\begin{figure}[h]
\begin{center}\includegraphics[%
%  width=60mm]{thesis/ah/Fig2.eps}\end{center}
 width=60mm]{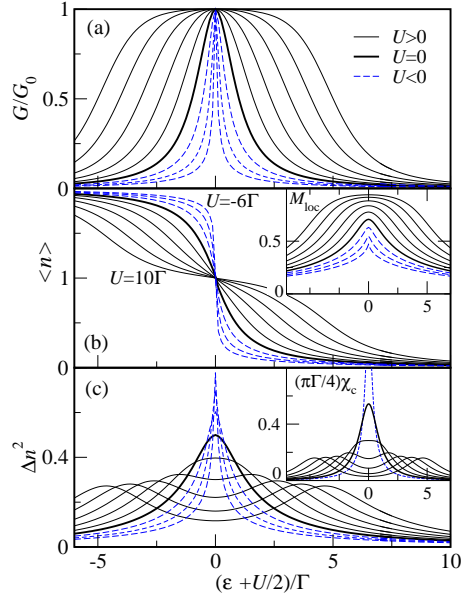}\end{center}
\caption{\label{cap:Fig2} (a) Conductance for the Anderson model with
  $-6\Gamma\leq U\leq10\Gamma$ in increments of $2\Gamma$ (full lines
  for $U>0$, dashed lines for $U<0$ and a thicker full line for
  $U=0$). (b) Local occupancy $n$ and local moment
  $M_{\mathrm{{loc}}}$ (inset). (c) Charge fluctuations $\Delta
  n^{2}=2n-n^{2}-M_{\mathrm{{loc}}}^{2}$.  Inset: renormalized charge
  susceptibility $(\pi\Gamma/4)\chi_{c}$. }
\end{figure}

The results presented here are discussed in more detail in
\cite{mravlje05}.  The results for repulsive $U$ as a function of gate
voltage $\epsilon +U/2$ are presented in Fig.~\ref{cap:Fig2} with full
lines. On the top panel the conductance, in the middle panel the
average charge and on the bottom panel the fluctuations of charge are
plotted. In the Kondo regime, where the average charge $\langle n
\rangle \sim 1$, the conductance is enhanced towards the unitary limit
$G \to G_0$. The maximal conductance is given by the quantum of
conductance $G_0 = 2e^2/h$ ($h$ is the Planck's constant, $e$ the
electron charge) and corresponds to the unitary transmission
\cite{datta95, houten96}. Actually, the average charge and conductance
are related by the Friedel sum rule
\begin{equation}
  G=G_{0}\sin^{2}{{\pi} \over {2}}n.
\end{equation}
The Friedel sum rule in this form holds for  single-impurity parity
symmetric models. The generalization to non-symmetric case
is possible \cite{mravlje06}.%  and is discussed later.  

The fingerprint of the Kondo physics is also the appearance of the
local moment $M_{\mathrm{{loc}}}=\bigl<
\left(n_{\uparrow}-n_{\downarrow}\right)^{2}\bigr>^{1/2}$, presented
in the inset of Fig.~\ref{cap:Fig2}(b) and the suppression of the
charge fluctuations $\Delta n^{2}=\bigl<\left(n-\langle n \rangle
\right)^{2}\bigr>$, Fig.~\ref{cap:Fig2}(c). In the inset the
corresponding charge susceptibility, $\chi_{c}=-\partial
n/\partial\epsilon$ is given. In agreement with the
fluctuation-dissipation theorem, the charge fluctuations are similar
to the charge susceptibility, $\Delta n^{2}\sim(\pi\Gamma/4)\chi_{c}$.
Strictly, $\langle n^{2}\rangle$ is given with the integral of the
imaginary part of the dynamic charge susceptibility,
$\chi_{c}^{''}(\omega)$, therefore the relation to static $\chi_{c}$
is only qualitative.

In Fig.~\ref{cap:Fig2}(a) the conductance for various $U<0$ is
presented with dashed lines. The first observation is a narrowing of
the conductance curve and the corresponding enhanced charge
fluctuations {[}Fig.~\ref{cap:Fig2}(c){]}, consistent with a sharp
transition in the local occupation and a suppression of the local
moment, Fig.~\ref{cap:Fig2}(b). For increasing  $|U|$, the charge
susceptibility diverges and overshoots the charge fluctuations.

For general values of parameters, i.e., for moderate $\Omega$, the
problem with EP coupling cannot be mapped onto the Anderson model.
However, the behavior is still to the largest extent determined by
$U_\mathrm{eff}$, and similar to the above discussed results,
Fig.~\ref{cap:Fig2}. For example, the result of the Schrieffer-Wolff
transformation is now an anisotropic Kondo model \cite{costi98}, where
the anisotropy stems from the fact that the phonon displacement
couples only to the $z$-component of the isospin $T_z$.

\begin{figure}[h]
\begin{center}\includegraphics[%
%  width=55mm]{novi_2_fon.eps}\end{center}
  width=60mm]{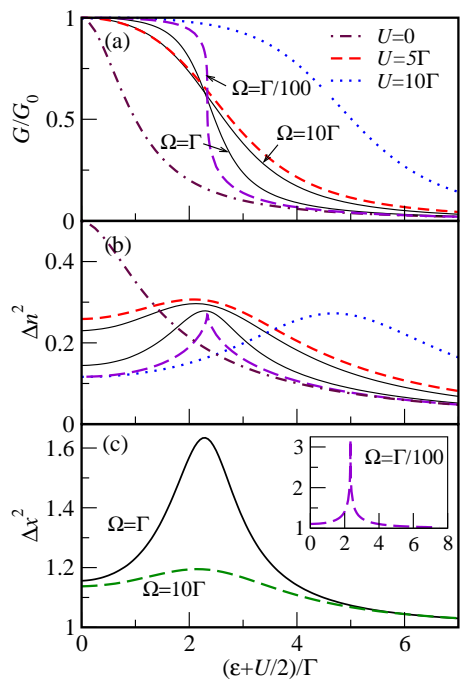}\end{center}
\caption{\label{cap:Fig4} A fixed $U=10\Gamma$ and
  $U_{\mathrm{eff}}=5\Gamma$ with for $\Omega=\Gamma$,
  $\Omega_{2}=10\Gamma$ and $\Omega_{2}=10\Gamma$. Also plotted are
  the results for a bare Anderson model with $U=10\Gamma$, $U=5\Gamma$
  and $U=0$ (dotted, short-dashed and dashed-dotted, respectively).
  (a) Conductance, (b) occupation fluctuations and (c) deformation
  fluctuations. In the inset, the deformation fluctuations for a
  softer mode are shown.}
\end{figure}

In addition to the renormalization of $U$ now also the hybridization is 
renormalized as shown
on Fig.~\ref{cap:Fig4}, where the results for bare $U=5\Gamma$ case
are compared to the $U=10\Gamma, U_\mathrm{eff} =5\Gamma$ case for
$\Omega=10\Gamma$, $\Omega=\Gamma$ and $\Omega=\Gamma/100$. The
smaller the $\Omega$, the sharper the jump in the conductance
corresponding to an enhanced effective mass due to the larger effect
of $e^{-\frac{M}{\Omega} (a-a^\dagger)}$.

The results for very soft phonons
$\Omega=\Gamma/100$ can also be understood in an alternative manner. In
the Kondo regime the conductance is close to the bare
Anderson model result with $U=10\Gamma$.  In the mixed valence regime
the curve is much steeper, due to a strong renormalization
\cite{schonhammer84} of the hopping. In the
empty-orbital regime the conductance approaches the result obtained
with a doubly reduced electron-electron interaction
\begin{equation} U_\mathrm{eff}=U-{{4M^{2}} \over \Omega},
\end{equation}
 which can be understood as follows. First
the oscillator displacement is shifted,
$x\rightarrow\tilde{x}+2\lambda$ thus the Hamiltonian is transformed
into 
\begin{equation} \tilde{H}=\left(\epsilon+2\lambda
M\right)n+\tilde{x}\left[M\left(n-1\right)+\Omega\lambda\right]+
\Omega\tilde{a}^{\dagger}\tilde{a}+...,
\end{equation} where
$\lambda=-M\left(n-1\right)/\Omega$, with vanishing transformed
displacement. This Hamiltonian can be solved with trial wave functions
with no phonons. The renormalized local energies are then
$\epsilon+2M^{2}/\Omega$, $\epsilon$, and
$\epsilon-2M^{2}/\Omega$ for $n=0,1,2$, respectively. The shifts
of $\epsilon$ where $n=0,\,2$ in turn correspond to reduced
$\tilde{U}=U-4M^{2}/\Omega$ and to $\tilde{U}=U$ for $n=1$.

\subsection{Softening of the phonon mode}
Phase transitions  are ubiquitously related to the
instability of the symmetry restoring modes. In ferroelectrics, for
instance, the  para$\to$ferro  phase  transition will occur when the
unstable phonon is frozen-in to one of the equivalent configurations
\cite{blinc74, anderson84}. As the temperature is tuned towards the transition,
$T\to T_c$ the related static temperature-dependent susceptibility
will diverge, $\chi(T)= C/(T-T_c)$. According to the Kramers-Kronig
relation
\begin{equation}
  \chi(T)=\chi'(0,T)=\frac{2}{\pi} \int_0^\infty
  {{\chi''(\omega',T)} \over {\omega'}} d\omega'
\end{equation} 
this will occur when the dissipative imaginary part of the susceptibility
$\chi''(\omega)$ has a peak at low frequencies. As
poles of $\chi''(\omega)$ indicate the normal modes of the system, the
frequency of the normal mode $\omega_0$ should vanish at the
transition.

\begin{figure}
\begin{center}\includegraphics[%
%  width=50mm]{thesis/softening_ah.eps}\end{center}
  width=55mm]{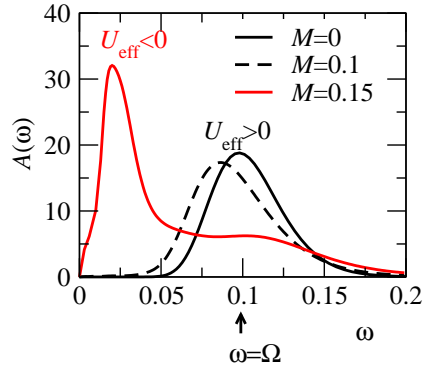}\end{center}
\caption{\label{cap:prop} The displacement-displacement spectral
  function for $\Omega/\Gamma=5$, $U/\Gamma=15$, $M/\Gamma=0,5,7.5$.}
\end{figure}

There is a remarkable analogy to this behavior in the
strong-coupling regime of the Anderson-Holstein model (although there
is no phase transition; we are dealing with a single degree of freedom
here). As shown in the inset of Fig.~\ref{cap:Fig2}(c), the charge
susceptibility $\partial n / \partial \epsilon$ diverges for large
$M$. The charge-charge correlation function should also increase there
according to the fluctuation-dissipation theorem. Due to the Holstein
coupling of charge to the displacement it seems plausible that the
phonon correlation function should also be influenced. Indeed, in the
Anderson-Holstein model the charge-charge and the
displacement-displacement correlation functions are directly related
\begin{equation}
D(\omega) = D_0(\omega) + M^2 D_0(\omega) \ll(n-1), (n-1)\gg_\omega
D_0(\omega), \label{eq:phononspec}
\end{equation}
as can easily be proved by considering the equation of motion. The
phonon propagator must thus develop a low frequency component. The
phonon mode is softened as $M$ grows large. On Fig.~\ref{cap:prop} the
NRG results for imaginary part of the phonon
propagator \begin{equation} \mathcal{A}(\omega)=-\frac{1}{\pi}
  \mathrm{Im} \ll x,x \gg_\omega  =-\frac{1}{\pi}\mathrm{Im}
  \int_{0}^{\infty} (-i) \langle[x(t),x(0)]\rangle e^{i \omega t}
  dt \end{equation} are plotted.
The oscillations which occur at frequency $\Omega$ for the uncoupled
oscillator become with increasing electron phonon coupling softer and
their characteristic frequency diminishes. The spectral functions are
broadened on the logarithmic scale as in \cite{mravlje08}. 
\begin{figure}
\begin{center}\includegraphics[%
%  width=50mm]{thesis/all_for_thesis.eps}\end{center}
  width=55mm]{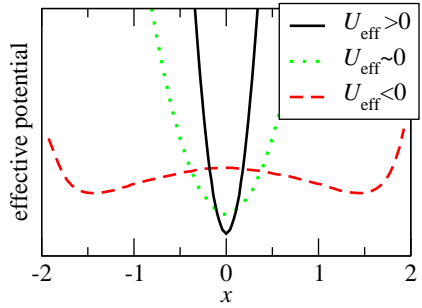}\end{center}
\caption{\label{cap:eff_pot} Effective potential for
  $U/\Gamma=7.5,\Omega/\Gamma=2.5$, and $M/\Omega=1.15,1.25,1.35$.}
\end{figure}

The softening can also be related to the change in the shape of the
effective potential the oscillator experiences due to the coupling to
the electrons. The effective potential can be extracted using the SG
method as explained in \cite{mravlje08}. The results are shown in
Fig.~\ref{cap:eff_pot}. We see that when the sign of the effective
repulsion is changed, the oscillator potential will evolve to a
double well form. The displacements of large magnitude (corresponding
to displaced oscillator transformation for states of double and zero
occupancy) will be preferred. The low frequency component of the
propagator corresponds to slow fluctuations of the oscillator
between the degenerate minima of the effective potential, the
high-frequency component corresponds to fast oscillations within
the wells.

\section{Oscillations with respect to the leads}
\begin{figure}
\begin{center}
  \includegraphics[width=45mm, keepaspectratio]{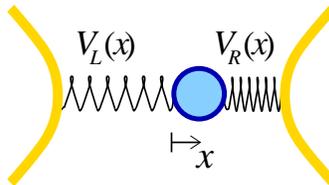}
\end{center}
\caption{\label{fig_cm_sketch} (Color online) Schematic plot of the model
  device.}
\end{figure}

We now turn to the case where the molecule oscillates between the
electrodes. We model the system with the Hamiltonian,
Eq.~(\ref{eq:hami}) for $M=0$, but with phonon-assisted hopping
induced by the displacement dependent tunneling matrix elements
$V_{k\alpha} \to V_{k\alpha}(x)$, as schematicaly presented on
Fig.~\ref{fig_cm_sketch}.

The full Hamiltonian thus reads, 
\begin{equation} 
H=\sum_{k\alpha} \epsilon_k n_{k\alpha} +
  \sum_{k\sigma}\left[ V_{k\alpha} (x) c_{k\alpha\sigma}^\dagger d_\sigma + h.c. \right]
  + \epsilon n + Un_\uparrow n_\downarrow + \Omega a^\dagger a.
\label{eq:hami2}
\end{equation}
From now on, we are here interested in the particle-hole symmetric
point $\epsilon=-U/2$, where the molecule is on average singly
occupied.  We use $U=15\Gamma$. Other details of calculation can be
found in \cite{mravlje08}. Again, it is practical to define even and
odd combinations of states in the electrodes, and in this basis the
tunneling part of the Hamiltonian reads,
\eb   V_e(x)\widehat{v}_e + V_o(x) \widehat{v}_o\ee
where  
\eb V_{e,o} (x)=\frac{V_L(x)\pm V_R(x)}{\sqrt{2}}, \ee
%$v_\gamma = \sum_{k\sigma} c^\dagger_{k\gamma\sigma}$ for
%$\gamma=e,o$, 
modulate the tunneling to even and odd channels. Hybridization
operators are $\widehat{v}_{\alpha}= f_{\sigma\alpha}^\dagger d_\sigma
+ h.c.$ for $\alpha=e,o$, respectively, where $f_e$ is the even
combination of electrode orbitals and $f_o$ is the odd combination of
electrode orbitals. Note that \eb\label{eq:ineq}|V_e(x)| >|V_o(x)|\ee
if $V_{L,R}(x)$ are both positive or both negative for all $x$.
\begin{figure}
\begin{center}
\includegraphics[width=80mm, keepaspectratio]{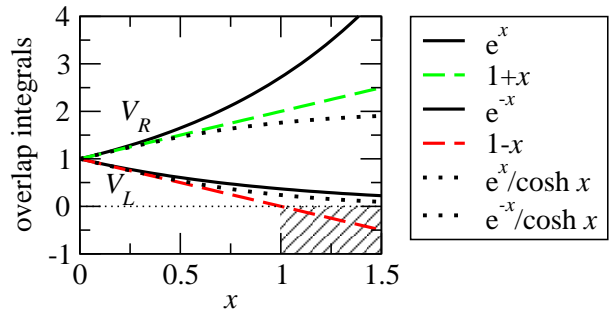}
\end{center}
\caption{\label{oscil_Fig_func} (Color online) Various forms of the
  tunneling-modulation.  The unphysical regime of LM where the
  tunneling starts to increase with increasing distance to the
  electrode is indicated by dashing.}
\end{figure}

\subsection{Two-channel Kondo model}
The odd channel is coupled to the molecule only due to the asymmetric
modulation of tunneling. For example, in the linear approximation
$V_{L,R}(x)=V(1 \mp g x )$ the even channel is coupled to the molecule
directly and the odd channel is coupled to the molecule via a term
proportional to $g x$. Unlike in the Anderson-Holstein model, the
attempts to eliminate the coupling to phonons using a variant of
Lang-Firsov transformation fail. Note that the coupling
to phonons as considered in this Section does not affect the effective
repulsion but it affects the hybridization and therefore the Kondo temperature
can be enhanced \cite{mravlje08}.

As a consequence of the coupling the molecular orbital to two channels the
low-energy behavior is that of the two-channel Kondo (2CK) model
\cite{nozieres80,balseiro06}.  The screening of the spin occurs in the
channel with the larger coupling constant. If the couplings match, an
overscreened, i.e., a genuine 2CK problem with a non-Fermi liquid
behaviour results. For a linearized model to be introduced below such
a fixed point has indeed been found at an isolated value of the
electron-phonon  coupling with simulations based on numerical
renormalization group \cite{mravlje07, mravlje08, silva09}.

\subsection{Overlap integrals}
The calculations are performed using several functional forms of
  $V_\alpha(x)$ depicted in Fig.~\ref{oscil_Fig_func}.
In a realistic experimental situation the tunneling between the
molecule and the tip of an electrode will be saturated at small
distances and it will progressively decrease with increasing distance of
the molecule from the electrode.  The precise functional dependence
of overlap integrals will in general depend on details of the molecule
and the tips of the electrodes, but the overall behavior should be as
shown in Fig.~\ref{oscil_Fig_func}(a) with dotted line.

\subsubsection {Linearized modulation }
The simplest form of overlap integrals is obtained by the expansion to
lowest order in displacement resulting in linear modulation (LM) \eb
V_{L,R} (x)=V\left[1\mp(gx +\zeta)\right]. \ee The tunneling matrix
element, constant $V$ for $g=0$, is linearly modulated by displacement
for $g > 0$. We assume the system is almost inversion symmetric. A
small $\zeta \geq 0$ is the magnitude of the symmetry breaking
perturbation. In the symmetrized basis the overlap integrals take on
the following form \eb\label{eq:linmod} V_e=\sqrt{2} V, \; \;
V_o=\sqrt{2}V (gx + \zeta).\ee Note that Eq.~(\ref{eq:linmod}) does
not satisfy the requirement Eq.~(\ref{eq:ineq}) for $gx> 1-\zeta$,
because  the overlap to the left electrode becomes negative and
its absolute value starts to increase with increasing $x$ (dashed
region in Fig.~\ref{oscil_Fig_func}).

\subsubsection{Regularized modulation}
A more realistic approximation to overlap integrals is the exponential
dependence on displacement but it breaks down at small distance to the
electrodes as discussed exhaustively in \cite{mravlje08}. The modulation should therefore 
at large  displacements be regularized and 
for the rest of this paper we use
\begin{equation} V_{L,R}(x)=V \left[ \exp(\mp gx)/\cosh(gx) \mp \zeta\right], \label{vtanh}
\end{equation}
or in the symmetrized basis \eb V_e=\sqrt{2} V,\: \: V_o =\sqrt{2} V
\left[\tanh(gx)+\zeta\right].\ee The inequality Eq.~(\ref{eq:ineq}) is
satisfied (the 2CK fixed point is thus inaccessible) and the
normalization with the cosh function ensures that the hybridization
saturates at small distances to the electrodes.

\begin{figure}
\begin{center}
\includegraphics[width=53mm, keepaspectratio]{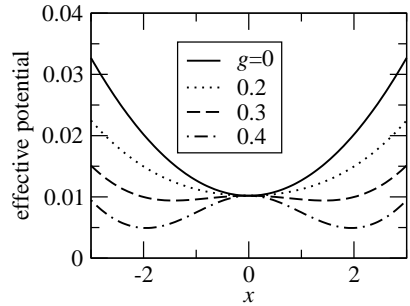}
\end{center}
\caption{\label{Fig_pot_simple}  Semi-classical
  estimate of effective oscillator potential for tanh
  modulation. Parameters $\Omega=0.01$, $\Gamma=0.02$ are in units of
  $D$ (half-width of the band). }
\end{figure}

\subsection{Effective potential}
At $U=0$ and replacing operators $a,x$ by real valued quantities, the
model Eq.~\ref{eq:hami2} is solvable exactly and the energy of the
ground state as a function of $x$ provides the estimate of the
effective oscillator potential. This simple estimate agrees
qualitatively with the results of more elaborate methods
\cite{mravlje08}. We plot the results on Fig.~\ref{Fig_pot_simple}. 

Initialy harmonic potential softens with increasing $g$ and at a
certain point a double well potential develops.  The softening thus
occurs similarly as in the case of Anderson-Holstein model but here it
is related to the dynamical breaking of inversion symmetry
\cite{balseiro06}. Due to the softening, the instability towards
perturbations breaking the symmetry (degeneracy between the two minima
of the double-well potential) can be expected. On the mean field level
\cite{mravlje06}, the instability is indeed seen as a tendency towards
an asymmetric ground state with large average $x$ in systems {\it
  with} inversion symmetry.

\section{Numerical results}
\begin{figure}
\begin{center}
  \includegraphics[width=90mm, keepaspectratio]{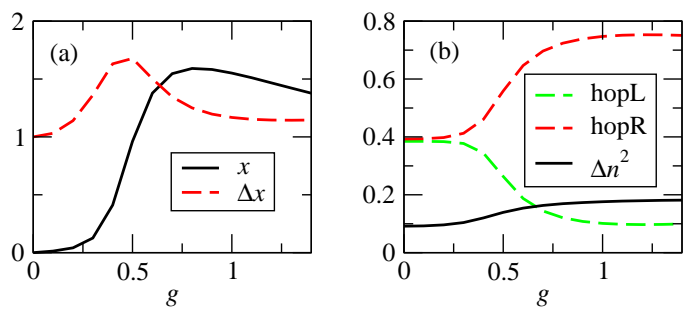}
\end{center}
\caption{\label{Fig_th} (Color online) (a) Displacement and
  displacement fluctuations. (b) Average hopping to left and
  right; fluctuations of charge.}
\end{figure}

The development of the double well potential induces fluctuations of
displacement and its influence can also be seen in the NRG results of
static quantities shown on Fig.~\ref{Fig_th} for
$U=0.3,\Gamma=0.02,\Omega=0.01$ In these results an inversion symmetry
breaking perturbation of strength $\zeta=0.01$ is included.

The average displacement presented on Fig.~\ref{Fig_th}(a) increases
as the electron-phonon coupling is increased. The fluctuations of
displacement initially increase then they dimininsh, as the oscillator
gets trapped in the lower of the two-well potential (also this
behavior is discussed in more detail in \cite{mravlje08}). At large
electron-phonon coupling also the average displacement starts to
diminish. This happens because for tanh-form of the hybridization for large $g$
the hybridization is maximal already for small displacemnts and
therefore the system can minimize the elastic energy without cost in
the kinetic energy.

On Fig.~\ref{Fig_th}(b) also the electronic expectation values are
shown. At large $g$, the molecule is near the right lead as signified
by increased hopping to the right. The total hybridization and,
correspondingly, the charge fluctuations are also increased there. The
Kondo temperature is increased, but the conductance
is suppressed due to the asymmetric configuration  \cite{mravlje08}.

\begin{figure}
\begin{center}
\includegraphics[width=60mm, keepaspectratio]{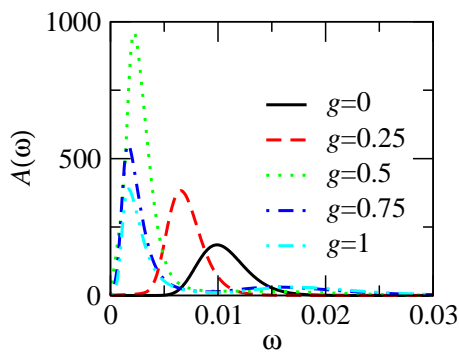}
\end{center}
\caption{\label{Fig_th_spec} (Color online) Displacement spectral
  functions for parameters as in Fig.~\ref{Fig_th}.}
\end{figure}

The softening of the potential is seen also in the displacement
spectral function shown on Fig.~\ref{Fig_th_spec}. Similar to the
behavior discussed in the previous chapter, the frequency of vibration
diminishes with increasing $g$, because the confining potential is
softened. At large $g$ the molecule is trapped to the lower of the two
wells, the oscillations between the two-wells become unfavourable and
the spectral weight is again transferred to high frequencies
corresponding to oscillations within the lower of the two wells.

\section{Conclusion}
We invoked Anderson model coupled to phonons to describe the transport
through the molecule coupled to molecular vibrations. While the
influence the MV exhibit on electron depends on the details of the
coupling, the coupling to electrons tends to soften the MV. In the
strong coupling regime of the Anderson-Holstein model ($U_\mathrm{eff}
< 0$), a perturbation of the orbital energy drives the system from
the particle-hole symmetric point characterized with zero average
displacement of the oscillator. Likewise, in the Anderson model with
asymmetricaly modulated hybridization, a perturbation of the
left-right symmetry results in a state with large average displacement
(the molecule is attracted to one of the electrodes).

\subsection{Discussion}
In measurements of conductance through molecular junctions the
side-peaks pertaining to the excitation/annihilation of vibrational
quanta are clearly discerned at a finite bias. The influence of
phonons in the equilibrium at a small bias is less investigated,
because it does not appear to affect the measured conductance
significantly. Why is this so?

The answer is that systems with electron-phonon (EP) coupling to
internal molecular modes can usually be reformulated in a manner that
refers to the EP coupling only by a redefinition (renormalization) of
the original (bare) parameters, which themselves are not,
unfortunately, known to start with.  For instance, such a parameter is
the repulsion, which is effectively diminished due to the influence of
the EP coupling. But because the repulsion is diminished already for a
decoupled molecule any attempt to discern the effects of the Holstein
phonon by comparing to the data for isolated molecules will likely
prove in vain.

If one was interested in discriminating the effects of the coupling to
phonons nevertheless, a convenient quantity to look at would be the
frequency dependence of the local charge susceptibility. In the regime
of reduced repulsion due to the EP coupling, the charge would be
susceptible to the oscillations of gate voltage only below the phonon
frequency. In order to investigate this experimentally, one would need
to be able to measure the time dependence of molecular charge. While
this currently seems a formidable task, we note that in quantum dots
(QDs) the time-resolved measurements of charge using quantum-point
contacts have already been demonstrated, e.g. in \cite{petta05}.

In near future, there is more hope to discern the effects of the EP
coupling to the contacts. However, even qualitative effects -- such as
the breaking of the particle-hole (PH) symmetry, which occurs when
the breathing modes couple to the electron charge and modulate the
tunneling at the same time -- can be dominated by non-perfectly
symmetric contacts. But still, measuring the linear conductance to a
better precision and comparing the data for rigid molecules to
the data for softer molecules could unravel the influence of the
breathing oscillations and the type of their coupling to the electron
transport in the equilibrium.

The case of molecules oscillating between the electrodes is a bit
different. One particular effect of the EP coupling is
obvious. Imagine a break junction with a single molecule bridging the
two contacts. In principle, one can slowly increase the tension
between the contacts, until they separate. Any tiny perturbation of the
parity will choose a contact to which  the bridging molecule will be
attracted. This is precisely the physics analyzed here: the tension
pulls the contacts apart and the separation sets the modulation of
tunneling by induced displacement of the molecule. With increasing tension
also the modulation increases. The potential confining the molecule to
the center is softened, and instantly, the molecule is attracted to one of the
contacts.

One might argue that the idea is rather to get as close to the
breaking point to observe the onset of the non-Fermi liquid (NFL)
fixed point. But because NFL fixed point corresponds to the case when
the molecule fluctuates between far left {\it and} far right, the
conductance is zero as is also if the molecule is far left {\it or}
far right.  It is worth calculating the temperature dependence of
conductance and hope for some anomalies due to the NFL formation
energy scale (for a very recent work in this direction see
\cite{silva09}), but we anticipate that it will be difficult to
distinguish the results from these corresponding to the displacement
of the molecule towards one of the contacts only.

\begin{figure}[h]
\begin{center}
  \includegraphics[width=40mm, keepaspectratio]{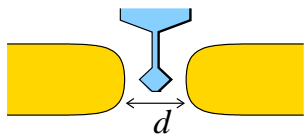}
\end{center}
\caption{\label{fig:nems}  Scheme of the proposed device.}
\end{figure}

On the other hand, one could observe the effects of the phonon-mode
softening directly in nanoelectromechanical suspended
cantilevers. Consider the setup, depicted in Fig.~\ref{fig:nems}.
The cantilever (blue) oscillates between the auxiliary contacts (gold).
 The frequency of oscillations can be detected independently,
 e.g.  by a quantum-point contact shifted perpendicularly with
 respect to the plane defined by the cantilever and the contacts
 \cite{poggio08}. We predict that the frequency of oscillations will
 decrease when the distance between the contacts $d$ is diminished
 provided that the device will operate in the regime of coherent
 tunneling. 

 It would be  interesting also to investigate the softening by
 suppressing it with the magnetic field. For magnetic fields above the
 Kondo temperature the spin at the orbital is frozen. According to the
 Pauli principle the charge fluctuations and the transport of
 electrons are blocked. Simultaneously, also the related kinetic
 energy gain vanishes and the softening is suppressed
 \cite{fabrizio09remark}.

%\bibliographystyle{spphys}
%\bibliography{yalta}

\end{document}